\documentclass[apjl]{emulateapj}
\usepackage[english]{babel}
\usepackage{amsmath}
\usepackage{amssymb}
\usepackage[dvips]{color}
\usepackage{natbib}

\def\la{\; \raise0.3ex\hbox{$<$\kern-0.75em\raise-1.1ex\hbox{$\sim$}}\;}
\def\ga{\;  \raise0.3ex\hbox{$>$\kern-0.75em\raise-1.1ex\hbox{$\sim$}}\;}

\begin{document}

\title[Anti-glitches within the standard scenario of pulsar glitches]
{Anti-glitches within the standard scenario of pulsar glitches}
%
%
\author{E. M. Kantor$^{1}$ and M. E. Gusakov$^{1,2}$}
\affiliation{
$^1$Ioffe Physical-Technical Institute of the Russian Academy of
Sciences,
Polytekhnicheskaya 26, 194021 St.-Petersburg, Russia
\\
$^2$St.-Petersburg State Polytechnical University,
Polytekhnicheskaya 29, 195251 St.-Petersburg, Russia
}


\date{Accepted 2014 xxxx. Received 2014 xxxx;
in original form 2014 xxxx}




%
\begin{abstract}
Recent observation of a sudden spin down, 
occurring on a timescale not exceeding two weeks,
of the magnetar 1E2259+586 
(see \citealt{archibald_etal_13}, 
where this event was 
dubbed
an `anti-glitch') 
has not still received any interpretation 
in terms of the standard scenario of pulsar glitches
proposed by \cite{AI75}. 
Motivated by this observation, 
here we present a toy model that allows, 
under certain conditions, 
for anti-glitches 
in neutron stars
within the standard approach.
\end{abstract}
%


\maketitle

\section{Introduction}
\label{Sec:Intro}
%
The generally accepted scenario of neutron star (NS) glitches, 
proposed by \cite{AI75}, 
assumes that sudden unpinning of a group of vortices from their pinning centers
results in 
an abrupt increase of the observed NS rotation frequency
--- a glitch. 
Here we present a toy model which demonstrates that, under certain conditions (in particular, 
for not too high neutron critical temperature in the outer core, see Sections \ref{Sec:Phys} and \ref{Sec:results} for details),
an opposite effect --- anti-glitch --- 
can take place due to avalanche-like unpinning of vortices.
Our model takes into account 
that superfluid 
density
depends on the relative velocities 
between the normal 
and superfluid NS components 
(hereafter the $\Delta V$-effect; 
see Section \ref{Sec:DeltaV}). 
When a group of vortices leaves the superfluid region, 
the velocity lag between the normal component
and pinned superfluid component decreases and, 
due to the $\Delta V$-effect, the mass of the superfluid fraction increases. 
Such a redistribution of mass 
between the normal and superfluid liquid components, 
which has been ignored so far,
can naturally lead to an anti-glitch for certain model parameters.
Although there are a number of models 
describing anti-glitches (see, e.g.,
\citealt{pbh12,pbh13,duncan13,lyutikov13,tong14,gr14,hg14}), 
the proposed effect allows one to explain the `sudden' spin down%
\footnote{The actual spin down timescale is not known but does not exceed two weeks. 
\cite{archibald_etal_13} dubbed this phenomenon an `anti-glitch'.}
of the magnetar 1E2259+586 (\citealt{archibald_etal_13}) 
within the generally accepted scenario of NS glitches.

\section{Superfluid density and the $\Delta V$-effect}
\label{Sec:DeltaV}

Here we 
introduce the notion of superfluid density $\rho_{s}$
and 
discuss how it can be affected 
by the relative motion of superfluid and normal currents induced in the system 
(see below for the definition of superfuid and normal currents).
Below we 
set
$\hbar=k_{\rm B}=1$.

Let us consider a non-relativistic degenerate Fermi-liquid composed of identical particles.
Assume that at a temperature $T$ less than some critical temperature $T_{c}$
they start to form Cooper pairs (become superfluid).
In what follows, we, for simplicity, 
assume that 
the particles pair in the spin-singlet $^1$S$_0$ state
and ignore the Fermi-liquid effects 
(in particular, we assume that the particle effective mass 
coincides with its bare mass $m$).

Generally, at $T<T_{c}$ 
it is instructive to think of
the superfluid liquid as consisting of two components,
the superfluid and normal ones.
The superfluid component, which has a density $\rho_s$, 
is associated with the Cooper-pair condensate, 
while the normal component with the density 
$\rho_{q}=\rho-\rho_{s}$ 
($\rho$ is the total liquid density)
is associated with the thermal (Bogoliubov) excitations 
(unpaired particles).
A notable property of any superfluid system is that 
these components can flow, without friction, 
with two distinct velocities (e.g., \citealt{ll80, khalatnikov00}).
As a consequence, the total mass-current density ${\pmb J}$ 
(or the momentum density ${\pmb P}$)
of the liquid can be presented as a sum of two terms,
\begin{equation}
{\pmb J}={\pmb P}=\rho_{s} {\pmb V}_{s}+\rho_{q} {\pmb V}_{q},
\label{J}
\end{equation}
where ${\pmb V}_{s}$ and ${\pmb V}_{q}$
are the velocities of the superfluid and normal components, respectively.
Note that the velocity ${\pmb V}_{s}$ 
is related to the momentum $2 {\pmb Q}$ of a Cooper pair 
by the expression ${\pmb V}_{s}={\pmb Q}/m$.
In the reference frame in which ${\pmb V}_{s}=0$, 
we have (see, e.g., \citealt{ll80, gh05})
\begin{equation}
{\pmb J}|_{{\pmb V}_{s}=0} = {\pmb P}|_{{\pmb V}_{s}=0} 
= \sum_{{\pmb p}, \sigma} \,  {\pmb p} \, f(E_{\pmb p}+{\pmb p} \, \Delta {\pmb V}), 
\label{PJ}
\end{equation}
where the summation goes over the Fermi momenta ${\pmb p}$ 
and spin states $\sigma$ of the Bogoliubov thermal excitations; 
$\Delta \pmb{V} \equiv {\pmb V}_{s}-\pmb{V}_{q}$; 
$f(x)=1/({\rm e}^{x/T}+1)$ is the Fermi-Dirac 
distribution function for thermal excitations; 
$E_{\pmb p}=\sqrt{v_{\rm F}^2 (|{\pmb p}|-p_{\rm F})^2 + \Delta^2}$ 
is their energy in the reference frame 
in which ${\pmb V}_{s}=0$;
$p_{\rm F}$ and $v_{\rm F}=p_{\rm F}/m$ 
are the particle's Fermi momentum and Fermi velocity, respectively.

Finally, $\Delta$ is the superfluid energy gap,
which generally depends on both $T$ and $|\Delta {\pmb V}|\equiv \Delta V$:
$\Delta=\Delta(T, \, \Delta V)$.
The fact that sufficiently large $\Delta V$ 
can affect the gap was emphasized by \cite{bardeen62} 
(see also \citealt{gk13,gj14}, 
where this effect was discussed in application to NSs).
In the absence of currents ($\Delta {\pmb V}=0$)
gap can be approximated as (\citealt{yls99})
\begin{equation}
\Delta(T,\, 0)=T \, \sqrt{1-\tau} (1.456 - 0.157/\sqrt{\tau}+1.764/\tau)
\label{gap0}
\end{equation}
($\tau \equiv T/T_c$) and decreases from the value 
$\Delta_{0}=1.764 \, T_{c}$ at $T=0$ 
to 0 at $T=T_{c}$.
The dependence of $\Delta$ on $\Delta V$ at $T=0$ is similar:
it decreases from $\Delta_{0}$ at $\Delta  V=0$ 
to 0 at $\Delta V\equiv \Delta V_{0}
={\rm e}\Delta_0/(2 p_{\rm F})$
(see \citealt{gk13} for details).
In an arbitrary frame 
equation (\ref{PJ}) can be rewritten as
\begin{equation}
{\pmb J} = {\pmb P} = \rho {\pmb V}_{s} 
+ \sum_{{\pmb p}, \sigma} \,  {\pmb p} \, f(E_{\pmb p}+{\pmb p} \, \Delta {\pmb V}).
\label{PJ2}
\end{equation}

Equations (\ref{J}) and (\ref{PJ2}) allow us to find expression for
the normal density $\rho_{q}$ and hence for the 
superfluid density $\rho_{s}=\rho-\rho_{q}$,
\begin{equation}
\rho_{q}= - \sum_{{\pmb p}, \sigma} \frac{{\pmb p}\Delta {\pmb V}}{\Delta V^2} \, 
f(E_{\pmb p}+{\pmb p} \, \Delta {\pmb V}).
\label{rhoq}
\end{equation}

As follows from this formula, $\rho_{q}$ and $\rho_{s}$
not only depend on $T$ but also on $\Delta V$ 
(this is the $\Delta V$-effect announced in the beginning of the section);
the latter dependence is especially pronounced at 
$p_{\rm F} \, \Delta V \sim 
\Delta(T,\, 0)$.
Usually, however, 
one considers 
a situation in which $p_{\rm F} \, \Delta V \ll \Delta(T,\, 0)$.
Then two simplifications can be made.
First, one can neglect the dependence of the gap on $\Delta V$,
$\Delta(T,\, \Delta V) \approx \Delta(T,\, 0)$
(\citealt{gk13}).
Second, one can expand the function $f(E_{\pmb p}+{\pmb p} \Delta {\pmb V})$ 
in equation (\ref{rhoq}) 
in Taylor series, retaining only the term linear in $\Delta {\pmb V}$.
The resulting expression for $\rho_{q}$ 
reduces then to the standard one (e.g., \citealt{ll80})
and is independent of $\Delta V$,
\begin{equation}
\rho_{q}= - \sum_{{\pmb p}, \sigma}
\frac{({\pmb p}\Delta {\pmb V})^2}{\Delta V^2} \, \frac{d f(E_{\pmb p})}{dE_{\pmb p}}.
\label{rhoq2}
\end{equation}

As it will be clear from the subsequent consideration, 
the condition $p_{\rm F} \, \Delta V \ll \Delta$ is not necessarily 
satisfied in NSs. 
Hence in what follows we will 
make use of the more general formula (\ref{rhoq}) 
rather than the standard equation (\ref{rhoq2}).
 
\begin{figure}
    \begin{center}
        \leavevmode
        \includegraphics[scale=0.7]{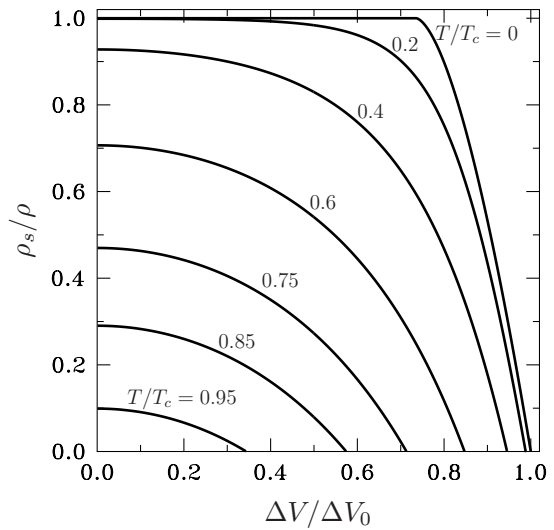}
    \end{center}
    \caption{Superfluid density $\rho_{s}$ in units of the total density $\rho$ as a function of the velocity lag $\Delta V$ normalized to $\Delta V_{0}$ for seven stellar temperatures.    
    }
    \label{Fig:frac}
\end{figure}

Figure \ref{Fig:frac} illustrates the dependence of 
$\rho_{s}=\rho-\rho_{q}$
on 
$\Delta V$ 
for seven stellar temperatures:
$T/T_{c}=0,\,0.2,\,0.4,\,0.6,\,0.75,\,0.85$, and $0.95$.
At $T/T_{c}\ll 1$, 
when $\Delta V$ is small, 
all particles are paired and $\rho_{s} \approx \rho$.
As the velocity lag $\Delta V$ becomes larger, 
the superfluid fraction decreases and
eventually disappears ($\rho_{s}=0$) 
at $\Delta V= \Delta V_{ 0}=e\Delta_0/(2p_{\rm F})$.

%

\section{How does the $\Delta V$-effect turn a glitch into an anti-glitch?}

The standard glitch scenario teaches us that when the 
crust
slows down, 
pinned superfluid stays 
rotating with a higher frequency $\Omega_{s}$, because vortices, which determine $\Omega_{s}$, cannot freely escape from the pinning region. 
(Note that by the `crust' we understand
not only the solid crust itself but 
also all the NS components rigidly coupled to it. 
They include non-superfluid and charged particles, 
neutron thermal Bogoliubov excitations,
as well as unpinned superfluid neutron component.)
At some moment a group of vortices unpins 
(the actual unpinning mechanism is unknown 
but is not important for us here) and transfers angular momentum to the crust.
In the standard scenario this event leads 
to decrease of $\Omega_{s}$ and, 
due to angular momentum conservation, to increase of $\Omega_{c}$. 
However, the standard consideration ignores the $\Delta V$-effect. How will $\Omega_{c}$ change if we account for it?
The answer follows from the angular momentum conservation,
\begin{equation}
I_{c0} \Omega_{c0}+I_{s0} \Omega_{s0}=
I_{c1} \Omega_{c1}+I_{s1} \Omega_{s1},
\label{0}
\end{equation}
where 
the subscripts $0$ and $1$ refer to the corresponding quantities 
before and after the glitch, respectively.
In equation (\ref{0}) $I_{s}$ is the moment of inertia 
of the pinned superfluid component;
$I_{c}=I-I_{s}$ is the moment of inertia of the remaining components;
$I$ is the total stellar moment of inertia.
In the non-relativistic limit $I_{s}$ is given by the integral
over the pinning region,
\begin{equation}
I_{s}=\int_{\rm pinning\, region} \rho_s\left(T,\,\Delta \Omega\, r\,{\rm sin}\theta
\right)r^2\,{\rm sin}^2 \theta \,\,{\rm d}V,
\label{Is}
\end{equation}
and depends on the rotation lag $\Delta \Omega$.
Here 
$\rho_{s}$ is the neutron superfluid density;
$r$ is the radial coordinate, 
$\theta$ is the polar angle, and ${\rm d}V$ is the volume element.

Taking into account that
the variation 
$\delta\Delta \Omega \equiv \Delta \Omega_1-\Delta \Omega_0=
(\Omega_{s1}-\Omega_{c1})-(\Omega_{s0}-\Omega_{c0})$ 
of the rotations lag in the glitch event is much smaller than $\Delta \Omega_0$, 
one can expand $I_{s}(\Delta \Omega)$ and $I_{c}(\Delta \Omega)$ 
in Taylor series near the point 
$\Delta \Omega = \Delta \Omega_0=\Omega_{s0}-\Omega_{c0}$
and rewrite equation (\ref{0}) in the form
\begin{eqnarray}
&&I_{c0} \Omega_{c0}+I_{s0} \Omega_{s0}=
\nonumber\\
&&[I_{c0}+I_{c}' (\delta \Omega_{s}-\delta \Omega_{c})] (\Omega_{c0}+\delta \Omega_{c})+\nonumber\\
&&[I_{s0}+I_{s}' (\delta \Omega_{s}-\delta \Omega_{c})] (\Omega_{s0}+\delta \Omega_{s}),
\label{1}
\end{eqnarray}
where
$I_{c}'=-I_{s}'=d I_{c}/d(\Delta \Omega)$;
$\delta \Omega_{s}=\Omega_{s1}-\Omega_{s0}$ and
$\delta \Omega_{c}=\Omega_{c1}-\Omega_{c0}$.
Equation (\ref{1}) yields 
\begin{equation}
\delta \Omega_{c}=-\frac{I_{s0}-I_{c}'\Delta \Omega_0}{I_{c0}+I_{c}'\Delta \Omega_0}\delta \Omega_{s}.
\label{main}
\end{equation}
This formula reduces to the standard result, 
$\delta \Omega_{c}=-I_{s0}/I_{c0}\,\delta \Omega_{s}$, 
if one sets $I_{c}'=0$.
Because $I_{c}' =-I_{s}'>0$ 
(superfluid density $\rho_{s}$ 
decreases with increasing $\Delta \Omega$), 
$\Delta \Omega_0=\Omega_{s0}-\Omega_{c0}>0$, 
and $\delta \Omega_{s}=\Omega_{s1}-\Omega_{s0}<0$,
it follows from equation (\ref{main}) that
we shall observe 
an abrupt pulsar spin down (an anti-glitch; $\delta \Omega_{c}<0$) if
$I_{c}' \Delta \Omega_0 >I_{s0}$ or, equivalently,
\begin{equation}
 |I_{s}'| \Delta \Omega_0 >I_{s0}.
 \label{cond}
\end{equation}
Thus, when anti-glitch occurs both the pinned superfluid and the 
rest of the star decelerate ($\delta \Omega_{c},\,\delta \Omega_{s} <0$), 
while the moment of inertia redistributes 
(via formation of additional Cooper pairs)
to satisfy the angular momentum conservation.

\section{Physics input}
\label{Sec:Phys}

As follows from equation (\ref{cond}), 
for anti-glitch to occur
it is necessary to have a relatively large 
$|I'_{s}|$. 
In other words, the superfluid neutron density
$\rho_{s}(T, \, \Delta \Omega \, r \, {\rm sin \theta})$
in the pinning region should be quite sensitive to 
a variation of $\Delta \Omega$ [see equation (\ref{Is})].
As discussed in Section 2, for that it is necessary
to have
 $p_{\rm F} \, \Delta V \sim 
\Delta(T,\, 0)$
or $p_{\rm F} \, \Delta \Omega \, r \, 
{\rm sin} \theta \sim 
\Delta(T,\, 0)$
[here and below the quantities 
$p_{\rm F}$, $\Delta(T,\, 0)$, $\Delta V$, $T_{c}$ 
etc.\ refer to neutrons].
From this estimate one obtains a typical 
value $\Delta \Omega_{\rm typ}$ of the rotation lag $\Delta \Omega$, 
at which one can expect to have an anti-glitch,
\begin{equation}
\Delta \Omega_{\rm typ} \sim 1 \left[ \frac{\Delta(T,\, 0)}{10^8 \, {\rm K}} \right] 
\left( \frac{n_0}{n} \right)^{1/3} 
\left( \frac{10^6 \, {\rm cm}}{r} \right) \,\,\,
{\rm rad} \,\, {\rm s}^{-1},
\label{typ}
\end{equation}
where $n_0=0.16$~fm$^{-3}$ is the nucleon number density in atomic nuclei;
$n=p_{\rm F}^3/(3 \pi^2)$ is the neutron number density.
Generally, $\Delta \Omega_{\rm typ}$ increases with $T_c$ 
(see equation \ref{gap0}).

The condition (\ref{typ}) can hardly be satisfied in the crust, 
because $T_{c}$ there is larger than 
$10^9$~K 
everywhere except for the narrow regions at the slopes 
of the neutron critical temperature profile $T_{c}(\rho)$. 
The corresponding values of $\Delta \Omega_{\rm typ}$ 
are much higher than possible frequency lags sustained by the vortex pinning 
(see \citealt{link14,spgh14}). 
But this condition is very likely to be met in the core.

\begin{figure}
    \begin{center}
        \leavevmode
        \includegraphics[scale=0.7]{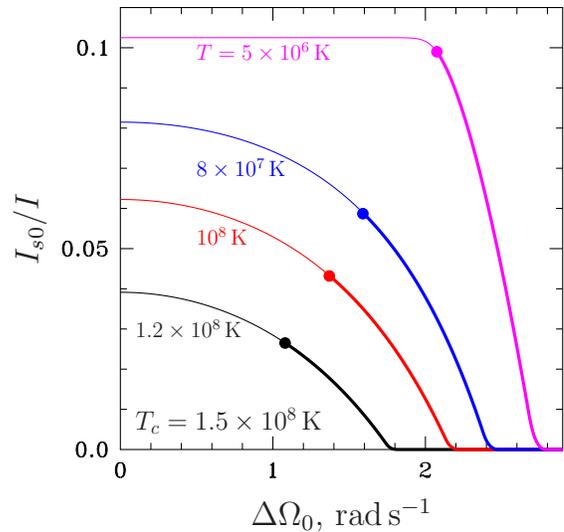}
    \end{center}
    \caption{Moment of inertia of the pinned superfluid $I_{s0}$ (in units of $I$) versus rotation lag, $\Delta \Omega_0$, for four values of the stellar temperature. 
For each temperature filled circle indicates a minimum value of $\Delta \Omega_0$ 
required to transform a glitch to an anti-glitch. 
At smaller $\Delta \Omega_0$ (thin lines) we observe glitches, 
at higher $\Delta \Omega_0$ (thick lines) --- anti-glitches.  
    }
    \label{Fig:Is}
    \vskip 0.15cm
\end{figure}

Recently \cite{chamel12} has demonstrated that entrainment in the crust 
can be strong.
If correct, this result indicates
that the crust is probably not enough to explain pulsar glitches 
so that the core superfluid may be responsible for glitches as well 
(\citealt{aghe12,chamel13}; see, however, \citealt{sgfn14, pfh14}). 
It is generally accepted (e.g., \citealt{bpp69}) that
protons in the core form type II superconductor 
which harbors magnetic field confined to flux tubes 
(for a discussion of other possibilities see 
\citealt{link03, jones06, cz07, ag08}
and references therein). 
Neutron vortices pin to magnetic flux tubes in the same way 
as they pin to nuclei in the crust. 
Purely poloidal configuration of the magnetic field 
cannot immobilize vortices so that they freely escape 
from the core as NS slows down. 
On the opposite, 
toroidal component efficiently prevents 
vortices from moving outwards (\citealt{sa09,ga14}). 
A number of numerical simulations 
in non-superfluid NSs (\citealt{braithwaite09,cr13}) 
and in superconducting NSs (\citealt{lander12,lander13,lander14}) 
show that toroidal component of the magnetic field, 
localized in outer layers of the NS core, 
is necessary for stability of the magnetic field configuration.
These simulations demonstrate that toroidal field, 
which can be noticeably higher than the surface magnetic field, 
is confined within the equatorial belt of the width $\sim 0.1R$ 
($R$ is the stellar radius) 
at a distance $r \sim 0.8R$ from the center 
(\citealt{lander12,lander13,lander14,cr13}). 
For definiteness, in our numerical 
calculations we assume that the pinning region, 
which coincides with the localization region of the toroidal field,
has the form of a torus with coordinates $r \in [0.75 R, \, 0.85 R]$,
$\theta \in [\pi/2-\pi/12, \, \pi/2+\pi/12]$,
$\phi \in [0, \, 2 \pi]$, where $\phi$ is the azimuthal angle.
The moment of inertia pinned 
to such 
a region
is about $I_{s}\sim 0.1 I$ for $\Delta V=0$ and $T=0$
(see Figure \ref{Fig:Is} and a similar estimate of \citealt{ga14}).

The rotation lag $\Delta \Omega_0$,
at which glitch/anti-glitch occurs, is uncertain.
Clearly, it cannot exceed the maximum value 
of the critical lag $\Delta \Omega_{\rm cr}$ 
in the pinning region,
required to unpin vortices from flux tubes by the Magnus force.
A typical energy of vortex pinning 
to flux tubes is of the order of $100\,\rm MeV$ 
[see \citealt{link14} and formula (A13) of \citealt{ruderman98}].
It corresponds to 
$\Delta \Omega_{\rm cr}\sim 0.1 B_{12}^{1/2}\,\rm rad\, s^{-1}$ 
(a more refined estimate can be found in \citealt{link14}).  
In case of magnetars we obtain $\Delta \Omega_{\rm cr}\sim (1-3) \,\rm rad\, s^{-1}$
for the magnetic field in the core $B \sim 10^{14}-10^{15}$~G. 
Below $\Delta \Omega_0$ 
will be treated as a free parameter of our toy model.

%

The value of 
$T_{\rm c}$
in the core is also uncertain 
and varies 
from $10^7$~K (\citealt{schwenk04})
to $10^9-10^{10}$~K (e.g., \citealt{baldo98}).
Note that, to calculate the gap
the latter authors used bare neutron-neutron interactions
and ignored medium polarization effects,
which can substantially overestimate $T_{\rm c}$ (\citealt{gps14}).
For illustration, in our calculations 
we choose $T_{c}=1.5\times 10^8\,\rm K$ in the 
pinning region ($T_c$ in the inner core can be larger).
This value does not contradict the results of microscopic calculations
and, e.g., is close to $T_{\rm c}$ reported by \cite{Lombardo13}.
It also agrees with 
the predictions of minimal cooling scenario 
(see, e.g., model a2 in figure 12 of \citealt{page13}).

\section{Results}
\label{Sec:results}

\begin{figure}[t!]
    \begin{center}
        \leavevmode
        \includegraphics[scale=0.7]{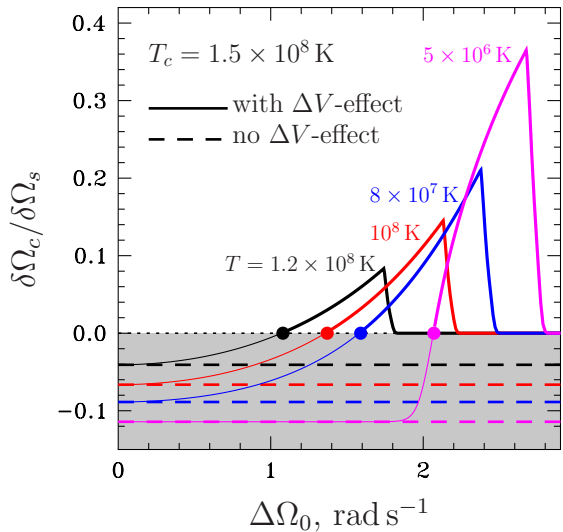}
    \end{center}
    \caption{Relative variation of the observed rotation frequency 
    versus initial rotation lag $\Delta \Omega_0$.
Dashed lines are calculated ignoring the $\Delta V$-effect. An anti-glitch corresponds to $\delta \Omega_{c}/\delta \Omega_{s}>0$ (unshaded region). Other notations are the same as in Figure \ref{Fig:Is}.
    }
    \label{Fig:dOmega}
    \vskip 0.2cm
\end{figure}

Using input parameters from 
Section 4,
we analyzed 
whether anti-glitches are possible 
if we allow for the $\Delta V$-effect.
Our results
are shown 
in Figures \ref{Fig:Is} and \ref{Fig:dOmega}. 
To plot the figures we employed APR equation of state (\citealt{apr98}) 
and considered a NS with the mass $M=1.4 M_\odot$.

Figure \ref{Fig:Is} presents the normalized moment of inertia of the pinned superfluid $I_{s0}$, 
calculated with equation (\ref{Is}), as a function of the rotation lag, $\Delta \Omega_{0}$. 
The function $I_{s0}(\Delta \Omega_0)$ 
is plotted for four typical (\citealt{kkpy14,pons13}) values of magnetar temperature, 
$T=1.2\times 10^8\,\rm K$ (black lines online), $T=10^8\,\rm K$ (red lines online), 
$T=8\times 10^7\,\rm K$ (blue lines online), and $T=5\times 10^6\,\rm K$ (magenta lines online). 
At small rotation lags 
$|I_{s}'|$ 
is too small to meet the condition (\ref{cond}) (see thin lines in the figure),
but it increases with $\Delta \Omega_0$.
Then, at some value of $\Delta \Omega_0$ (marked with filled circles) 
$|I_{s}'| \Delta \Omega_0$ and $I_{s0}$ become equal to one another
and at higher $\Delta \Omega_0$ the inequality (\ref{cond}) is always hold 
(then a vortex avalanche leads to an anti-glitch; thick lines in the figure).

Figure \ref{Fig:dOmega} presents 
the ratio $\delta \Omega_{c}/\delta \Omega_{s}$ [see equation (\ref{main})]
versus 
$\Delta \Omega_0$.
We remind the reader that $\delta \Omega_{c}$ is the observed 
rotation frequency jump, while $\delta \Omega_{s}$
is the frequency jump of the pinned superfluid component.
In the standard glitch scenario 
$\delta \Omega_s$ is negative and depends 
on the number of unpinned vortices and on 
the place in the star where they repin.
It is thus a poorly constrained parameter. 

%

The curves are plotted for the same
set of temperatures
as in Figure \ref{Fig:Is}.
When $\Delta \Omega_0$ is small the $\Delta V$-effect is negligible 
and solid lines almost coincide with the corresponding dashed lines, 
which are calculated ignoring this effect. 
As $\Delta \Omega_0$ increases, 
$\delta \Omega_{c}/\delta \Omega_{s}$ also increases and
eventually 
reaches 0 (this moment is shown by filled circles in the figure).
At larger rotation lags
$\delta \Omega_{c}$ becomes negative 
($\delta \Omega_{c}/\delta \Omega_{s}$ is positive);
then vortex avalanche leads to an anti-glitch,
which generally has a similar size $|\delta \Omega_{c}|$ 
as glitches at the same $\delta \Omega_{s}$.
Note that, for higher temperatures $\Delta(T,\, 0)$ is smaller
and a lower rotation lag is required to produce an anti-glitch [see estimate (\ref{typ})]. 

Further increase of $\Delta \Omega_0$ 
leads to a gradual shrinking of the pinning region
(due to transformation of the superfluid matter 
to normal matter with growing $\Delta \Omega_0$).
This leads to a rapid decrease of $|I_{s}'|$ 
(see Figure \ref{Fig:Is}) and hence to a sharp decrease of 
$\delta \Omega_{c}/\delta \Omega_{s}$.
Finally, when the superfluidity is completely destroyed in the whole pinning region,
neither glitches nor anti-glitches are possible 
($\delta \Omega_{c}/\delta \Omega_{s} = 0$).

As follows from Figure \ref{Fig:dOmega},
 anti-glitches 
can be produced 
for $\Delta \Omega_0 \ga (1-2)$~rad~s$^{-1}$,
in agreement with the 
estimate (\ref{typ}).
Such values of $\Delta \Omega_0$ are comparable to the critical rotation 
lag $\Delta \Omega_{\rm cr}$, estimated in Section 4 (recall that $\Delta \Omega_0$ cannot
exceed $\Delta \Omega_{\rm cr}$). 
Thus, the standard 
glitch 
scenario 
can
account for anti-glitches.
Note, however, that this conclusion is sensitive to the assumed value of $T_c$
in the pinning region. 
For example, for $T_c \sim 10^9$~K (e.g., \citealt{baldo98}) one has 
$\Delta \Omega_0 \sim \Delta \Omega_{\rm typ} \sim 10$~rad s$^{-1}$;
for such $T_c$ anti-glitches will hardly occur in our model 
unless $\Delta \Omega_{\rm cr}$ is substantially larger 
for some reason.

\section{Discussion and conclusion}

Here we propose a toy model that allows us to 
describe an anti-glitch
within the standard scenario of pulsar glitches formulated by \cite{AI75}. The main feature of our model is account for the so-called $\Delta V$-effect 
--- the dependence of the 
superfluid density on the relative velocity of normal and superfluid components (see Section \ref{Sec:DeltaV}). 

We predict that magnetars are the most promising anti-glitching objects. 
High magnetic field of magnetars provides strong pinning of vortices to flux tubes 
in the outer core, 
which leads to a large rotation
lag between the normal and superfluid components. 
As we showed, such a rotation lag
may be sufficient 
to transform a glitch to an anti-glitch. 

Could similar mechanism produce anti-glitches in the {\it crust}
of a magnetar or an ordinary pulsar?
Most probably not, because 
the critical rotation lag $\Delta \Omega_{\rm cr}$
(see Section 4)
seems to be noticeably smaller (\citealt{link14,spgh14}) 
than the typical rotation lag 
$\Delta \Omega_{\rm typ}$ 
[see equation (\ref{typ})],
which is needed to affect $\rho_{s}$.
There are two reasons for that.
First,
the pinning force per unit length 
is weaker for pinning to crust nuclei than for pinning 
to flux tubes in the core of a magnetar. 
Second, the rotation lag
($\sim \Delta \Omega_{\rm typ}$),
that can lead to an anti-glitch in the crust,
is severalfold larger than in the core, 
because the neutron critical temperature $T_{c}$ in the crust is higher. 
Due to these two factors it seems implausible 
that vortex avalanches in the crust can lead 
to a substantial redistribution of the stellar moment of inertia 
and thus to an anti-glitch.
However, if, due to some reason, 
vortex unpinning occurs exclusively in the 
narrow region 
where $T_{c}$ is substantially lower 
(i.e., on the slopes of the critical temperature profile), 
then anti-glitches could, in principle, be produced.

We summarize by concluding that: 
($i$) the same 
magnetar
can exhibit both glitches (due to vortex unpinning in the crust) and, 
under certain conditions, 
anti-glitches (due to vortex unpinning in the core) and
($ii$) it is not very likely that ordinary pulsars can demonstrate any anti-glitching activity.

\acknowledgements

This is an extended version of the contribution presented 
at the conference ``Physics of neutron stars --- 2014''
(July 28 -- August 1, 2014, St.-Petersburg, Russia).
We are very grateful to its participants, especially,
to 
Crist$\acute{{\rm o}}$bal Espinoza,
Bennett Link, 
and Andreas Reisenegger
for discussions and thoughtful comments.
This study was supported by the Russian Science Foundation 
(grant number 14-12-00316).

\bibliographystyle{apj}

\label{lastpage}

\end{document}